\documentclass[aps,preprint,superscriptaddress,nofootinbib]{revtex4}

\usepackage{color}
\usepackage{epsfig}
\usepackage{ulem}
\newcommand{\be}{\begin{equation}}
\newcommand{\ee}{\end{equation}}

\newcommand{\bea}{\begin{eqnarray}}
\newcommand{\eea}{\end{eqnarray}}
\newcommand{\beas}{\begin{eqnarray*}}
\newcommand{\eeas}{\end{eqnarray*}}
\newcommand{\n}{\nonumber}

\newcommand{\lam}{\lambda}

\newcommand{\ba}{\begin{array}}
\newcommand{\ea}{\end{array}}

\newcommand{\bg}{\bar{\gamma}}

\begin{document}

\title{Zero energy states of Dirac equation in $(2+1)$-dimensional curved spacetime}

\author{Choon-Lin Ho
}
\address{Department of Physics, Tamkang University, Tamsui
251, Taiwan, R.O.C.}

\author{Pinaki Roy}
\address{Atomic~Molecular~and~Optical~Physics~Research~Group, Advanced Institute of Materials Science, Ton Duc Thang University, Ho~Chi~Minh~City, Vietnam\\
Faculty of Applied Sciences, Ton Duc Thang University, Ho Chi Minh City, Vietnam}




\begin{abstract}
We consider Dirac equation in $(2+1)$ dimensional curved spacetime in the presence of a scalar potential. It is then shown that the zero energy states are degenerate and they can be obtained when the momentum $k_y$ in the $y$ direction satisfies certain constraints involving the mass parameter and the scalar potential coupling.
  
\end{abstract}



\maketitle

\section{ Introduction.} 

In recent years there have been a renewed interest in the study of lower dimensional Dirac equation. The main reason for this is that the motion of the quasiparticles in Dirac materials like graphene, fullerene, silicene, germanene, topological insulators etc. are described by the $(2+1)$ dimensional Dirac equation \cite{G1,G2,G3,W1}. In such materials it is a challenging problem to  confine the quasiparticles and it has been shown to be possible by using different types of magnetic fields \cite{m1,m2,m3,m4}.  On the other hand it is generally believed that because of Klein tunneling electric fields are not useful in confining the quasiparticles in graphene. However during the past few years it has been shown that using certain types of electric fields it is indeed possible to confine the quasiparticles in graphene and zero energy states can be obtained analytically for a number of electric field profiles \cite{z1,z2,z3,z4,z5,z6,z7,z8,z9,z10,z11,z12,z13,z14}.

It may be noted that graphene and other $2d$ materials may exist not only in flat form but also in different curved shapes and in such cases Dirac equation in curved space has to be used to study electronic or transport properties \cite{c1,c2,c3,c4,c5}. Furthermore, there has been an increasing interest in curved graphenes as they have been considered as possible analogue gravity media, namely, systems employed  to study general relativistic effects in laboratory \cite{AG1,gtr2,gtr3,gtr4}.

In this Letter our objective is to extend the analysis of zero energy states in flat spacetime \cite{z1,z2,z3,z4,z5,z6,z7,z8,z9,z10,z11,z12,z13,z14} to curved spacetime. More precisely we shall consider certain types of $(2+1)$ dimensional curved spaces and examine whether or not zero energy states of massive as well as massless Dirac equation can be obtained analytically in such spacetime in the presence of a scalar potential. It will be seen that depending on the nature of the curved space, zero energy states can be obtained analytically provided the parameters of the model satisfy certain constraints. Another interesting result is that for the zero energy states to exist the momentum $k_y$ in the $y$ direction can not assume arbitrary values but must be restricted in certain ranges. Finally it has been also been shown that these states are degenerate and the degeneracy has also been determined.

\section{The  Model}

Consider a $(2+1)$ dimensional curved spacetime with a metric $g_{\mu\nu}, \mu, \nu=0,1,2$. We use the Greek indices to refer to the curved spacetime coordinates.
The Dirac equation of a particle with charge $e$ and mass $M$ in this curved spacetimes with a minimal coupling to an electromagnetic potential $A_\mu$ is
\be
\left[i\gamma^\mu\left(\partial_\mu +\Gamma_\mu\right)-e\gamma^\mu\,A_\mu -M\right]\Psi\left(x,y,t\right)=0.
\label{Dirac1}
\ee
Here the Dirac matrices $\gamma^\mu$ satisfy
\be
\{\gamma^\mu, \gamma^\nu\}=g^{\mu\nu},
\ee
and
$\Gamma_\mu$ are the spinor affine connections required to give the covariant derivatives of a spinor.

In curved spacetimes, we need to  express the Dirac gamma matrices $\gamma^\mu$ in terms of those in a Minkowskian local inertial frame (the existence of which is made possible due to the equivalence principle).  This is  facilitated by the use of the vielbeins $e^a_\mu$, where the Latin indices refer to the local inertial frame.  The vielbeins  $e^a_\mu$ and their inverse $E^\mu_a$ are used to project tensors between the two frames, and they satisfy \cite{MD,KNA,Lima}
\bea
e^a_{~\mu} e^b_{~\nu} \eta_{ab}&=&g_{\mu\nu},\n\\
E_a^{~\mu} e_b^{~\nu} g_{\mu\nu} &=& \eta_{ab}.
\eea
where $\eta_{ab}={\rm\ diag} (1, -1, -1)$  is the Minkowskian metric.  

Let $\bg^a$ be the Dirac gamma matrices in the Minkowskian spacetime, which satisfy
\be
\{\bg^a, \bg^b\}=2\eta^{ab}.
\ee
Then $\gamma^\mu$ and the spinor affine connections $\Gamma^\mu$ are given by
\bea
\gamma^\mu &=& \bg^a E_a^{~\mu},\n\\
\Gamma^\mu &=& \frac18\omega_{ab\mu} [\bg^a, \bg^b].
\eea
Here $\omega^a_{~b\mu}$ is the spin connection given by the vielbeins  $e^a_{~\nu}$ and the covariant derivatives of their inverses $E^\nu_{~b;\mu}$:
\bea
\omega^a_{~b\mu} &=&e^a_{~\nu} E^\nu_{~b;\mu}\n\\
&=& e^a_{~\nu}\left(\partial_\mu E_b^{~\nu} + \Gamma^\nu_{\sigma \mu} E_b^{~\sigma}\right),
\eea 
where  $\Gamma^\nu_{\sigma \mu}$ are the Christoffel symbols.

In this note we consider the spacetime defined by the line element
\be
ds^2=\Omega^2(x)\left(dt^2-dx^2\right)-dy^2,
\ee
where $\Omega(x)$ is a function of $x$. The $(1+1)$-dimensional case $ds^2=\Omega^2(x)\left(dt^2-dx^2\right)$ was considered in Ref [32,33], with emphasis on quantum mechanics of a modified supersymmetric harmonic oscillator, and on quantum field theory, respectively.

With this metric, the vielbeins and their inverses are $e^a_{~\mu}={\rm\ diag} (\Omega (x), \Omega (x), 1)$ and $E_a^{~\mu}={\rm\ diag} (1/\Omega (x), 1/\Omega (x), 1)$.  The Christoffel symbols and the spin connections are the same as those obtained in Ref. 33,
\be
\Gamma^0_{01}=\Gamma^0_{10}=\Gamma^1_{00}=\Gamma^1_{11}=\frac{\Omega^\prime}{\Omega}, ~~{\rm \ others}=0,
\ee
and 
\be
\omega^0_{~10}=\omega^1_{~00}=\frac{\Omega^\prime}{\Omega}, ~~{\rm \ others}=0.
\ee
Here $\Omega^\prime =d\Omega (x)/dx$.

The Dirac gamma matrices $\gamma^\mu$ are
\be
\gamma^0=\frac{\bg^0}{\Omega}, ~\gamma^1=\frac{\bg^1}{\Omega}, ~\gamma^2=\bg^2,
\ee
and the spinor affine connections 
\be
\Gamma_0=\frac{\Omega^\prime}{4\Omega}[\bg^0, \bg^1], ~\Gamma_1=\Gamma_2=0.
\ee

Let us now consider the case of only a pure scalar potential , i.e., $e A_\mu(x,y, t) =(V(x), 0,0)$.  
The Dirac equation (\ref{Dirac1}) becomes
\be
\left\{i\left(\partial_0 +\frac{\Omega^\prime}{2\Omega}\bg^0\bg^1 \right)- V(x) + i\bg^0 \bg^1\partial_1
+ i\Omega\bg^0 \bg^2 \partial_2  - \Omega \bg^0 M\right\} \Psi =0.
\label{Dirac2}
\ee

\section{Zero energy states}

Exact analytical solutions of all energy values of Eq.\,(\ref{Dirac2}) are in general difficult to find.  It may however be possible to find the exact analytical zero energy state.  In recent years zero energy states are of considerable interest in condensed matter physics, particularly in systems related to graphene \cite{z1,z2,z3,z4,z5,z6,z7,z8,z9,z10,z11,z12,z13,z14}.

We restrict the space coordinates to be $x\in (-\infty, \infty)$ and $y\in [0, L]$.
We choose following representation of the gamma matrices in terms of the Pauli matrices
\be
\bg^0=\sigma_3, ~~ \bg^1=i\sigma_2, ~~\bg^2=i\sigma_1.
\ee
Also, we assume the ansatz
\be
\Psi(x,y,t)=\frac{1}{\sqrt{\Omega}} e^{-iEt+ik_y y}  \psi(x),
\label{wf}
\ee
where $\psi(x)$ is a two-component spinor, $E$ and $k_y$ are the energy and momentum in the $y$-direction.
Eq. (\ref{Dirac2}) then reduces to
\be
\left[-i\sigma_1\partial_1-\sigma_2\left(k_y \Omega(x)\right) +\sigma_3\left(M \Omega(x)\right) +V(x)-E\right]\psi(x)=0.
\label{Dirac3}
\ee

It is interesting to note that Eq.(\ref{Dirac3}) can be identified with that of a $(1+1)$ dimensional flat space Dirac oscillator with a position-dependent mass \cite{HR}.
Following the approach discussed in Ref. 35, we consider zero energy solutions of one class of potential, which is also proportional to $\Omega(x)$, i.e.,  $V(x)=k_v \Omega(x)$. 
Eq.(\ref{Dirac3}) with $E=0$ then becomes
\be
\partial_1\psi(x)=-\left(\sigma_3 k_y + \sigma_2 M+ i\sigma_1 k_v\right)\,W(x)\,\psi(x).
\ee
Now we assume the solution to have the form
\be
\psi(x)=\chi\,\phi(x),
\ee
where the spinor $\chi$ satisfies
\be
\left(\sigma_3 k_y + \sigma_2 M+ i\sigma_1 k_v\right)\chi=\lam\,\chi.
\ee
The eigenvalues are 
 $\lam_\sigma=\sigma\sqrt{k_y^2  + M^2  -k_v^2}, \sigma=\pm 1$, and the corresponding eigenfunctions are
 \bea\label{chi}
 \chi_\sigma\sim 
 \left(
 \begin{array}{c}
   1 \\
  i \frac{\lam_\sigma- k_y }{M-k_v}
  \end{array}
  \right).
 \eea
$\phi_\sigma(x)$ is solved from the equation
 \be
 \partial_x\phi_\sigma(x)=-\lam_\sigma \Omega(x)\phi_\sigma(x),
 \ee
 whose solution is
 \be\label{zero}
\phi_\sigma(x)\sim e^{-\lam_\sigma\int^x dx \Omega(x)}.
\ee
For the $E=0$ solution, one should choose $\lam_\sigma$ and $\Omega(x)$ such that $\Psi(x,y,t)$ in (\ref{wf})  is normalizable.

Now the presence of $\Omega(x)$ in the denominator of $\Psi(x,y,t)$ implies that $\Omega(x)$ must be nodeless in $(-\infty, \infty)$, and that $|\Omega(x)|\to 
\infty$ as $|x|\to \infty$.  Note that this behaviour of the function $\Omega(x)$ makes (\ref{zero}) non normalizable if $\lambda_\sigma$ is real. Thus an admissible zero state can exist only if $\lam_\sigma$ be pure imaginary or zero. This requires $k_v^2 \geq k_y^2  + M^2$ i.e., $\lam_\sigma=i\sigma \lam$, $\lam=\sqrt{k_v^2-k_y^2-M^2}\geq 0$. This implies that the momentum $k_y$ is sort of quantized in the sense it can only assume values in the interval $[-\sqrt{k_v^2-M^2},\sqrt{k_v^2-M^2}]$. It may also be pointed out that in the massless case the condition simply becomes $k_v^2\geq k_y^2$.  Furthermore, if we impose periodic boundary condition in the $y$-direction, then $k_y L=2\pi m, m=0, \pm 1, \pm 2, \ldots$. In this case above constraint becomes $-\frac{L\sqrt{k_v^2-M^2}}{2\pi}\leq m \leq \frac{L\sqrt{k_v^2-M^2}}{2\pi}$. Therefore the degeneracy of the zero energy state is given by $2\lfloor \frac{L\sqrt{k_v^2-M^2}}{2\pi}\rfloor +1$, where $\lfloor x \rfloor$ denotes the floor function. So finally the zero energy solutions are given by
\begin{equation}
\Psi_{\sigma}(x,y,t)=N_\sigma\frac{1}{\sqrt{\Omega(x)}}e^{ik_y y} \chi_{\sigma}\phi_{\sigma}(x),
\end{equation}
where $\chi_{\sigma}$ and $\phi_{\sigma}$ are given by (\ref{chi}) and (\ref{zero}) respectively, and $N_\sigma$ is a normalization constant.

We determine the normalization constant 
 $N_\sigma$ by the inner product defined over the $x^0={\rm\ constant}$ Cauchy hypersurface with metric $g_{ij}={\rm\ diag}(-\Omega(x), -1)$ ($i, j=1,2$)\cite{Par}
\be
\langle \Psi_\sigma | \Psi_\sigma\rangle=\int\,dxdy \sqrt{g_2}\,\rho (x,y,t),
\ee
where $g_2=det (g_{ij})=\Omega (x)$, and the probability density  $\rho$ is defined by
\be
\rho=\Psi^\dagger\bg^0\gamma^0\Psi=\frac{1}{\Omega(x)}\Psi^\dagger\Psi.
\ee
With $\Psi(x,y,t)$ given by (\ref{wf}), we have
\be
\rho(x)=\left[L \Omega(x)^2\int_{-\infty}^\infty dx\, \frac{1}{\Omega(x)}\right]^{-1}.
\ee

Since for zero energy states to exist the constraint on $\Omega(x)$ is rather mild there are many options  to choose it. We consider some of the possibilities given below  for the purpose of illustration ($c,\omega>0,  n=1,2,3,\ldots$):

1.  $\Omega (x)=(\omega x^{2n} +c)$;

2.  $\Omega (x)=\cosh^n  \alpha x$;

Let us consider Case 1 where for simplicity we have taken $c=1,\omega=1$. The normalization constants for $n=1,2,3$ are given by
\begin{equation}
	N_1=\sqrt{\frac{(k_v-M)}{2\pi L k_v}}, ~N_2=\sqrt{\frac{(k_v-M)}{\sqrt{2}\pi L k_v}},~ N_3=\sqrt{\frac{3(k_v-M)}{4\pi L k_v}},
\end{equation} 
and a plot of the corresponding normalized scaled probability densities $P(x)=\sqrt{g_2} L \rho(x)$ are given in Fig.\,1.
\begin{figure}[h]
	\includegraphics{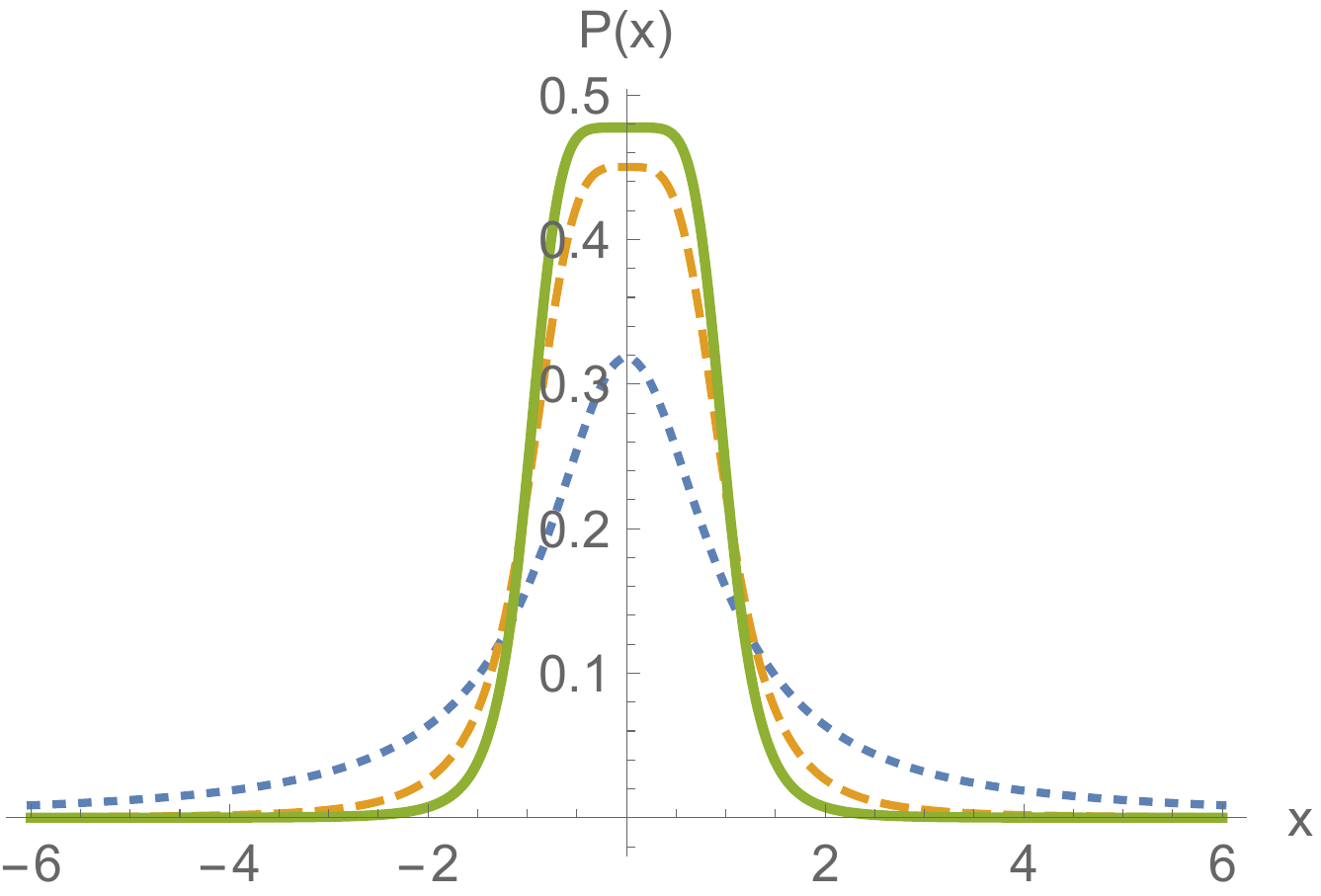}
	\caption{Normalised scaled probability densities $P(x)$ for $n=1$ (dotted curve), $n=2$ (Dot dashed curve) and $n=3$ (Solid curve).}
\end{figure}

\newpage
For Case 2,  the normalization constants can be obtained exactly and for $\alpha=1, n=1,2,3$ they are given by
\begin{equation}
	N_1=\sqrt{\frac{(k_v-M)}{2\pi L k_v}},~ N_2=\sqrt{\frac{(k_v-M)}{4 L k_v}},~ N_3=\sqrt{\frac{(k_v-M)}{\pi L k_v}}.
\end{equation}
The corresponding normalized scaled probability densities $P(x)$ are shown Fig.\,2.
\begin{figure}[h]
	\includegraphics{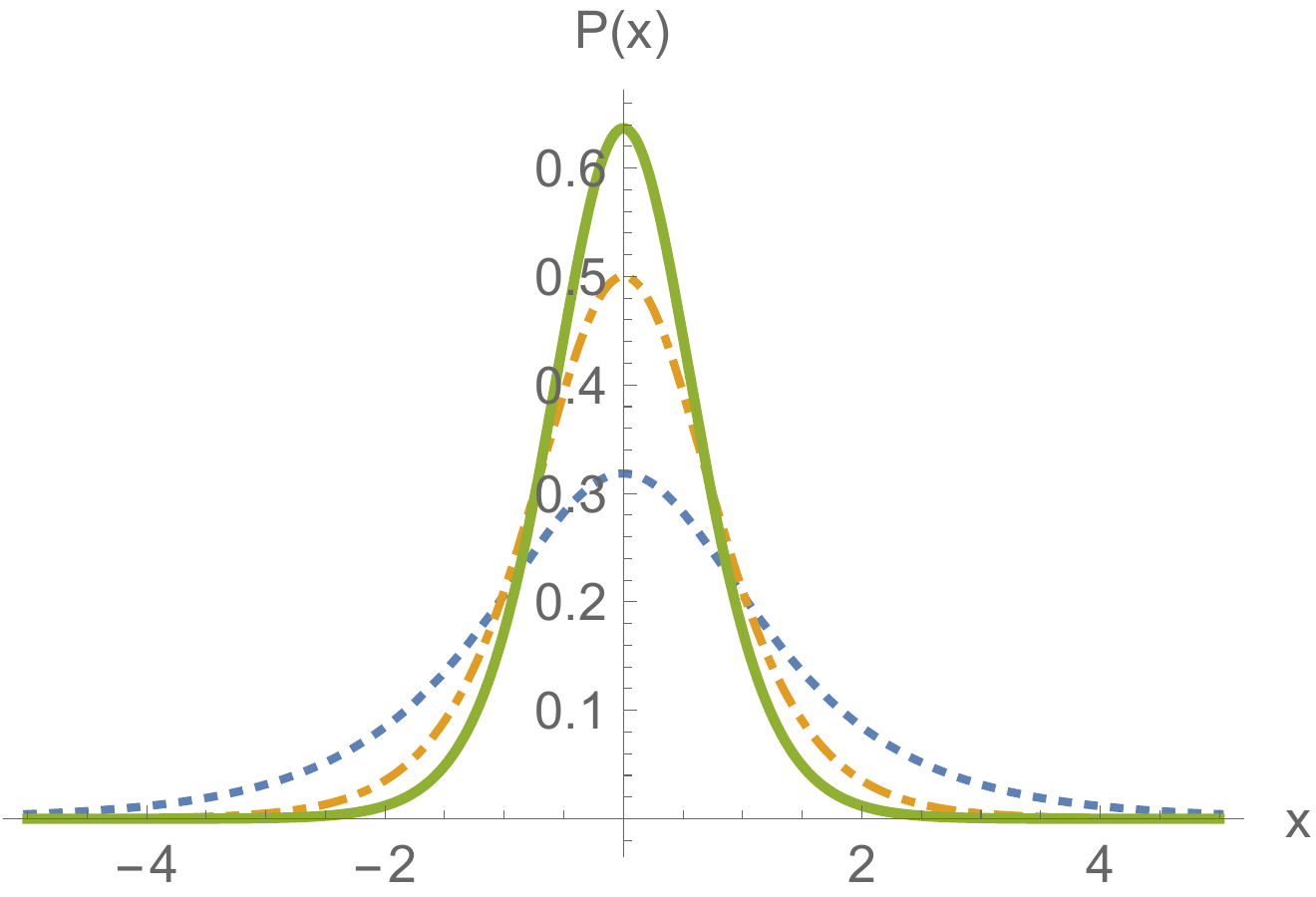}
	\caption{Normalised scaled probability densities $P(x)$ for $n=1$ (dotted curve), $n=2$ (Dot dashed curve) and $n=3$ (Solid curve).}
\end{figure}
\section*{Acknowledgments}

The work is supported in part by the Ministry of Science and Technology (MoST)
of the Republic of China under Grant MOST 110-2112-M-032-011. 


\end{document}